\documentclass[a4paper,11pt]{article}
\usepackage{graphicx,amssymb,amstext,amsmath,mathabx,mathtools,esvect,xparse}
\usepackage{color,wrapfig}
\usepackage{floatflt}
\usepackage{lipsum,ragged2e}

\usepackage{cite}

\usepackage{float}
\usepackage{esint}
\usepackage{subcaption}
\usepackage{cite}
\usepackage{mathtools}
\usepackage{commath}
\usepackage{hyperref}

\definecolor{cbl}{rgb}{0,0,1}                % bleu\

\newcommand{\bc}{\begin{center}}
\newcommand{\ec}{\end{center}}
\def\ba#1{\begin{array}{#1}\displaystyle}
\newcommand{\ea}{\end{array}}

\newcommand{\beq}{\begin{equation}}
\newcommand{\eeq}{\end{equation}}
\newcommand{\beqa}{\begin{eqnarray}}
\newcommand{\eeqa}{\end{eqnarray}}

\newcommand{\bi}{\begin{itemize}}
\newcommand{\ei}{\end{itemize}}
\newcommand{\bra}{\langle}
\newcommand{\ket}{\rangle}

\begin{document}
\begin{titlepage}
\vspace{0.2cm}
\begin{center}

{\large{\bf{Symmetry Resolved Entanglement with $U(1)$ Symmetry:\\ Some Closed Formulae for Excited States}}}

\vspace{0.8cm} 
{\large Olalla A. Castro-Alvaredo$^\heartsuit$ and Luc\'ia Santamar\'ia-Sanz$^\diamondsuit$}

\vspace{0.8cm}
{\small

$^{\heartsuit}$ Department of Mathematics, City, University of London, 10 Northampton Square EC1V 0HB, UK\\
\medskip

\hspace{-0.5cm}$^\diamondsuit$ Department of Applied Mathematics and Computer Science, University of Burgos, Hospital del Rey s/n, Burgos, 09001, Spain \\
}
\end{center}

\medskip
\medskip
\medskip
\medskip
In this work, we revisit a problem we addressed in previous publications with various collaborators, that is, the computation of the symmetry resolved entanglement entropies of zero-density excited states in infinite volume. The universal nature of the charged moments of these states has already been noted previously. Here, we investigate this problem further, by writing general formulae for the entropies of excited states consisting of an arbitrary number of subsets of identical excitations. When the initial state is written in terms of qubits with appropriate probabilistic coefficients, we find the final formulae to be of a combinatorial nature too. We analyse some of their features numerically and analytically and find that for qubit states consisting of particles of the same charge, the symmetry resolved entropies are independent of region size relative to system size, even if the number and configuration entropies are not. 
\noindent 
\medskip
\medskip
\medskip
\medskip

\noindent {\bfseries Keywords:} Quantum Entanglement, Symmetry Resolved Entanglement, Excited States, Qubits

\vfill
\noindent 
{$^\heartsuit$}o.castro-alvaredo@city.ac.uk\\
$^\diamondsuit$ lssanz@ubu.es

\hfill \today

\end{titlepage}
%\tableofcontents
\section{Introduction}
The study of entanglement measures in the context of low-dimensional quantum field theory (QFT) is a popular field of research within theoretical physics. Notably, the universal properties of the von Neumann entropy of critical models which allow for the characterisation of critical points have been known for a long time, both from analytical and numerical works \cite{CallanW94,HolzheyLW94,latorre1,Calabrese:2004eu,Latorre2,Jin,latorre3}. 
In more recent times, a lot of attention has been devoted to interrogating how symmetries and their breaking manifest on entanglement measures. For example, the {\it entanglement asymmetry} is a measure that quantifies the excess entanglement associated with symmetry breaking in a subsystem of a quantum system \cite{Ares:2022koq} and its study has been carried out for many critical and off-critical systems \cite{Capizzi:2020mio,Ares:2023kcz,Bertini:2023ysg,Capizzi:2023xaf,Ares:2023ggj}.

In this work we are interested in the case when an internal symmetry is present. Then, an entanglement measure known as {\it symmetry resolved entanglement entropy} (SREE) has been introduced which quantifies the contribution of each symmetry sector to the total entanglement entropy \cite{GS,german3}. The SREE and other similarly generalised entanglement measures have been extensively studied since 2018 and we have recently co-authored a review on the subject \cite{ourreview}. An important property of the SREE is that it (or at least a particular contribution to it) can be experimentally measured \cite{Islam,expSRE1,expSRE2,expSRE3,expSRE4,Lukin_2019}, which provides additional motivation for its study.  

In order to define the measure of interest we need to introduce some basic quantities. Let $|\Psi\ket$ be a pure state and let us define a bipartition of space into two complementary regions $A$ and $\bar{A}$ so that the Hilbert space of the theory $\mathcal{H}$ also decomposes into a direct product $\mathcal{H}_A \otimes \mathcal{H}_{\bar{A}}$. Then the reduced density matrix associated to subsystem $A$ is obtained by tracing out the degrees of freedom of subsystem $\bar{A}$ in
\beq
\rho_A=\mathrm{Tr}_{\bar{A}}(|\Psi\ket \bra \Psi|)\,,
\eeq
and the von Neumann and $n$th R\'enyi entropy of subsystem $A$ are defined as
\beq
S=-\mathrm{Tr}_A(\rho_A \log \rho_A)\quad \mathrm{and} \quad S_n=\frac{\log(\mathrm{Tr}_A \rho_A^n)}{1-n}\quad \mathrm{with}\quad S=\lim_{n\rightarrow 1} S_n\,,
\label{SS}
\eeq
where $\mathrm{Tr}_A \rho_A^n:={\mathcal{Z}}_n/{\mathcal{Z}}_1^n$ can be interpreted as the normalised partition function of a theory constructed from $n$ non-interacting replicas of the chosen model. 

 In the presence of an internal symmetry, if the symmetry operator $Q$ commutes with the hamiltonian of the system, its projection $Q_A$ onto subsystem $A$ satisfies $[Q_A, \rho_A]=0$. Thus, if $q$ is the eigenvalue of operator $Q_A$, then we can define ${\mathcal{Z}}_n(q)=\mathrm{Tr}_A(\rho_A^n \mathbb{P}(q))$ with $\mathbb{P}(q)$ the projector onto the charge $q$ symmetry sector, as the {\it symmetry resolved partition function}. The $n$th symmetry resolved R\'enyi entropy and the SREE of the sector of charge $q$ are then
\beq
S_n(q)=\frac{1}{1-n} \log\frac{{\mathcal{Z}}_n(q)}{{\mathcal{Z}}_1^n(q)}\quad \mathrm{and} \quad S(q)=\lim_{n\rightarrow 1} S_n(q)\,,
\label{sym}
\eeq
in analogy to (\ref{SS}). 
As proposed in \cite{GS} the partition functions ${\mathcal{Z}}_n(q)$ can be expressed in terms of their Fourier transforms, the {\it charged moments} $Z_n(\alpha)=\mathrm{Tr}_A(\rho_A^n e^{2\pi i\alpha Q_A})$ as
\beq
\mathcal{Z}_n(q)=\int_{-\frac{1}{2}}^{\frac{1}{2}} d\alpha \, Z_n(\alpha) e^{-2\pi i\alpha q}\,,
\label{moments}
\eeq
where we now specify the symmetry to be  $U(1)$. This will be the case of interest in this work. 

An interesting property is that the total entropy $S$ and the SREEs, $S(q)$, are related by the formula:
\beq 
S=\sum_q \left[p(q) S(q)- p(q) \log p(q)\right]\,,
\label{number}
\eeq 
where $p(q):=\mathcal{Z}_1(q)$ represents the probability of obtaining the value $q$ when measuring the charge. In the context of symmetry resolved measures, the first contribution is known as {\it configuration entropy} which, as we can see, is the sum of the SREEs weighted by the probability of the charge sector. The second contribution is the {\it number entropy} or {\it fluctuation entropy}, which is associated to fluctuations in the value of the charge of subsystem $A$. When the conserved quantity is particle number, 
the number entropy gives information about particle-number fluctuations between subsystems and is experimentally measurable \cite{Lukin_2019, Wiseman}. 

Starting from these basic definitions, SREEs have been computed and discussed for many classes of models and we refer to the review \cite{ourreview} for a comprehensive list of published related works. Since this is a short paper, we would like to focus our attention on those works that relate most directly to the present discussion. The motivation for this work goes back to results on the {\it excess entropy} of excited states, that is the difference between the entanglement entropy of an excited state and that of the ground state. In the case of critical systems or conformal field theory, and of low-lying excited states, this was studied in \cite{german1,german2}. Later on, gapped systems in the form of massive free quantum field theories and gapped quantum spin chains were studied and it was found that in a certain scaling limit and for certain kinds of excited states, the excess entropy takes an extremely simple and universal form. This form depends only on the ratio of entanglement regions' size to total system size, called $r$ and $1-r$, for a bipartition, and on the statistics of excitations. It does not depend on the masses, momenta, or energies of the excited quasiparticles. 

For excited states consisting of a finite number of excitations and large system and subsystem sizes, universal formulae for the excess entropy were obtained in \cite{excited,excited1,excited2,excited3}. The results were originally derived by employing the branch point twist field approach \cite{entropy} in free theories. However, it was shown in \cite{excited} for several examples, that the formulae should hold for interacting and even higher-dimensional theories, as long as excitations are localised. These results have been confirmed and then extended in many ways by later works, \cite{ali1, Zhang_2021a,Zhang1,Zhang2,Zhang3,Zhang4, Angel_Ramelli_2021} and also interpreted within a semiclassical picture \cite{Mussardo:2021gws,excitedAlba}. 

A particularly recent and interesting extension has been  to symmetry resolved measures in complex free theories with $U(1)$ symmetry \cite{ourPartI,ourPartII,ourPartIII}. The aim of this work is to generalise some of the results found in \cite{ourPartI}, particularly in relation to the {\it qubit picture}.
\medskip

Our work is organised as follows: In Section \ref{revision} we revisit the results of \cite{ourPartI}, in particular their derivation for qubit states. In Section \ref{generalise} we generalise those results to a state with any number of distinct and indistinguishable excitations. Distinct excitations may be so because their momenta, their symmetry charges or both, are distinct, but results are the same in all cases. We conclude in Section \ref{conclude}. 

\section{SREEs of Excited States: Some Known Results}
\label{revision}
As an approach to obtaining the excess entanglement of excited states in the scaling limit, the qubit picture was first used in \cite{excited, excited1}. It was realised that the same formulae for the excess entropy obtained from lengthy QFT computations could be derived much more simply by computing the entanglement associated with specific states consisting of a finite number of qubits. For example, the excess entanglement due to a single excitation above the ground state is the same as the bipartite entanglement of a state of two qubits 
given by
\beq 
|{\bf 1}\ket = \sqrt{r}\, |1\ket \otimes |0\ket + \sqrt{1-r} \, |0 \ket\otimes |1\ket\,,
\label{1par}
\eeq 
where the square of the coefficients, $r$ and $1-r$, is interpreted as the probability of finding the excitation in region $A$ and in its complement, respectively, and the state $|1\ket \otimes |0\ket$ represents one excitation in region $A$ and no excitation in its complement $\bar{A}$. The von Neumann entropy of such a state is simply $-r \log r - (1-r)\log(1-r)$ and the nth R\'enyi entropy is $(1-n)^{-1}\log(r^n+(1-r)^n)$. States containing any number of excitations in any region can be constructed similarly. For example, in a more complicated situation, with two identical and one distinct excitation present, we can write the corresponding state
\beqa 
|{\bf 1,2}\ket &=& \sqrt{r^3} |1 2\ket \otimes |00\ket+ \sqrt{2r^2(1-r)} |11\ket \otimes |01\ket +  \sqrt{r(1-r)^2} |10\ket \otimes |02\ket \nonumber\\
&& + \sqrt{r^2(1-r)} |0 2\ket \otimes |10\ket+ \sqrt{2r(1-r)^2} |01\ket \otimes |11\ket +  \sqrt{(1-r)^3} |00\ket \otimes |12\ket\,.
\eeqa 
and the associated R\'enyi entropy, denoted by $S_n^{1,2}(r)$ is the sum of the result for one excitation already given above plus the result for two identical excitations, that is, 
\beq 
S_n^{1,2}(r)=\frac{1}{1-n}\log(r^n+(1-r)^n)+ \frac{1}{1-n} \log(r^{2n}+2^n r^n (1-r)^n + (1-r)^{2n}).
\label{exam}
\eeq 
This construction is very easy to generalise given its obvious combinatorial nature. 
%Although computing the entanglement entropy of these states is much simpler than a QFT computation, it is important to remember that for qubits the formulae above give the full entanglement associated to a bipartition of the system whereas in a QFT they just give the extra entanglement due to the excitations, that is, there is a highly non-trivial contribution from the ground state which is subtracted in QFT.
In the case of symmetry resolved measures the result is slightly different, although related to the above. In this case, what is universal is not the excess entropy but the {\it ratio of charged moments} between the excited and the ground states. For the symmetry resolved R\'enyi entropies, let us call this ratio $M_n^\Psi(r;\alpha)$ for a generic state $|\Psi \ket$ and Fourier parameter $\alpha$. The most important formulae found for this ratio correspond to cases of distinct and non-distinct excitations.  For a state $|{\bf 1}^\epsilon \ket$ of a single particle excitation with $U(1)$ charge $\epsilon$ we have that 
\beq
M_n^{1^\epsilon}(r;\alpha)=e^{2\pi i \epsilon \alpha} r^n + (1-r)^n\,,
\label{una}
\eeq
whereas for a state of $k$ identical excitations of charge $\epsilon$ we have that
\beq
M_n^{k^\epsilon}(r;\alpha)= \sum_{j=0}^k [f_j^k(r)]^n e^{2\pi i j \epsilon \alpha}\,,
\label{for1}
\eeq
where $ f_j^k(r):={}_kC_{j} \, r^j (1-r)^{k-j}$ and ${}_kC_{j}=\frac{k!}{j!(k-j)!}$ the binomial coefficient. Formula (\ref{for1}) is in fact the building block that allows us to obtain the ratio of charged moments for any other configurations. For a generic state comprising $s$ groups of $k_i$ identical particles of charge $\epsilon_i$, denoted by $|{\bf k_1^{\epsilon_1} k_2^{\epsilon_2}\ldots k_s^{\epsilon_s}}\ket $, we will have 
\beq
M_n^{k_1^{\epsilon_1}\ldots k_s^{\epsilon_s}}(r;\alpha)=\prod_{i=1}^s M_n^{k_i^{\epsilon_i}}(r;\alpha)\,.
\label{general}
\eeq
For $\alpha=0$ these formulae reduce to those found in \cite{excited,excited1} for the partition function and would for instance allow us to reproduce the result (\ref{exam}) with $s=2$, $k_1=1$ and $k_2=2$.

In order to obtain the SREE from the formulae above it is necessary to isolate the charged moments of the excited state. For the state of one excitation, formula (\ref{una}) means that the ratio $M_n^{1^\epsilon}(r,q)$ is
\beq 
\frac{Z_n^{1^\epsilon}(r,\alpha)}{Z_n^0(r,\alpha)}= e^{2\pi i \alpha} r^n + (1-r)^n\,,
\eeq 
where we have introduced the $r$ dependence in the notation for the charged moments. 
We then have from the definition (\ref{moments}) that the partition functions are related as
\beqa
\mathcal{Z}_n^{1^\epsilon}(r,q)&=&\int_{-\frac{1}{2}}^{\frac{1}{2}}d\alpha \, Z_n^0(r,\alpha) e^{-2\pi i \alpha q} \left(e^{2\pi i \epsilon \alpha} r^n + (1-r)^n\right)\nonumber\\
&=& \mathcal{Z}_n^0(r,q-\epsilon)r^n+\mathcal{Z}_n^0(r,q) (1-r)^n\,,
\eeqa
where we also now make the $r$-dependence explicit. 
This kind of formula can be clearly generalised to other states, as we discuss later. 
It also implies that the SREEs of these kinds of excited states can be written in terms of the SREEs of the ground state. For instance, for the same state of one excitation considered above, the symmetry resolved R\'enyi entropies are
\beq
S^{1^\epsilon}_n(r,q)=\frac{1}{1-n}\log\frac{\mathcal{Z}^{1^\epsilon}_n(r,q)}{(\mathcal{Z}^{1^\epsilon}_1(r,q))^n}=
\frac{1}{1-n} \log\frac{
\mathcal{Z}_n^0(r,q-\epsilon) r^n +\mathcal{Z}_n^0(r,q)(1-r)^n}{(\mathcal{Z}_1^0(r,q-\epsilon) r +\mathcal{Z}_1^0(r,q)(1-r))^n}\,, 
\label{SnR1}
\eeq
Here $\mathcal{Z}^{1^\epsilon}_n(r,q)$ are the symmetry resolved partition functions in the state $|{\bf 1^\epsilon}\ket$ and $\mathcal{Z}^{0}_n(r, q)$ are those of the ground state. 

As before, the case of qubit states is simpler. For such states there is no notion of ground state. In other words, the charged moments $Z_n^0(r,\alpha)=1$ and therefore, the symmetry resolved partition functions are just
\beq
\mathcal{Z}_n^0(r,q) =\int_{-\frac{1}{2}}^{\frac{1}{2}} d\alpha \, e^{-2\pi i \alpha q}=\frac{\sin \pi q}{\pi q}=\delta_{q,0}\qquad {\rm for}\quad q\in \mathbb{Z}\,,
\label{funk}
\eeq
that is, they just implement the projection over the sector of total charge $q$.  The symmetry resolved R\'enyi entropy (\ref{SnR1}) further simplifies to:
\beq
S^{1^\epsilon}_n(r,q)= \frac{1}{1-n} \log\left[\frac{\delta_{q,\epsilon} r^n + \delta_{q,0} (1-r)^n}{\left(\delta_{q,\epsilon} r + \delta_{q,0} (1-r)\right)^n}\right]\,,
\label{s1}
\eeq
and for the case of $k$ identical excitations of charge $\epsilon$ we can employ (\ref{for1}) to obtain 
\beq
S^{k^\epsilon}_n(r,q)=\frac{1}{1-n}\log\frac{\sum_{j=0}^k \left[f_j^k(r)\right]^n \delta_{q,\epsilon j}}{\left[ \sum_{j=0}^k f_j^k(r) \delta_{q,\epsilon j}\right]^n}\,,
\label{sk}
\eeq
It is easy to see that once we fix the change $q=\epsilon j$ for some integer $j$, the formulae (\ref{s1}) and (\ref{sk}) give the value zero for the SREE. This is because for identical excitations symmetry resolution does not add any extra information. In fact, in this case, the von Neumann entanglement entropy equals the number entropy  (see formula (\ref{number})). As already observed in \cite{ourPartI} non-trivial results are only obtained when the state contains at least some distinct excitations. The aim of this work is to investigate and interpret the results obtained for those cases systematically. 

\section{Symmetry Resolved Partition Functions and Entropies}
\label{generalise}
Consider again the formula (\ref{for1}) and (\ref{general}). We have then that the symmetry resolved partition function for a generic state 
$|{\bf k_1^{\epsilon_1} k_2^{\epsilon_2}\ldots k_s^{\epsilon_s}}\ket$ can be obtained as
\beqa 
\mathcal{Z}_n^{k_1^{\epsilon_1}\ldots k_s^{\epsilon_s}}(r,q)&=& \sum_{j_1=0}^{k_1}\sum_{j_2=0}^{k_2}\cdots \sum_{j_s=0}^{k_s} \int_{-\frac{1}{2}}^{\frac{1}{2}} d\alpha \, Z_n^0(r,\alpha) \left[\prod_{\ell=1}^s f_{j_\ell}^{k_\ell}(r) \right]^n e^{2\pi i (\sum_{\ell=1}^s j_\ell \epsilon_\ell-q) \alpha}\nonumber\\
&=& \sum_{j_1=0}^{k_1}\sum_{j_2=0}^{k_2}\cdots \sum_{j_s=0}^{k_s}  \, \mathcal{Z}_n^0(r,q-\sum_{\ell=1}^s j_\ell \epsilon_\ell) \left[\prod_{\ell=1}^s f_{j_\ell}^{k_\ell}(r) \right]^n\,. 
\label{genZn}
\eeqa
From this general formula the symmetry resolved R\'enyi and von Neumann entropies can be obtained from the definitions (\ref{sym}). In particular, the von Neumann entropy can be written as
\beq 
S^{k_1^{\epsilon_1}\ldots k_s^{\epsilon_s}}(r,q)=-\lim_{n\rightarrow 1} \frac{\partial_n \mathcal{Z}_n^{k_1^{\epsilon_1}\ldots k_s^{\epsilon_s}}(r,q)}{\mathcal{Z}_n^{k_1^{\epsilon_1}\ldots k_s^{\epsilon_s}}(r,q)}+\log \mathcal{Z}_1^{k_1^{\epsilon_1}\ldots k_s^{\epsilon_s}}(r,q)\,,
\label{sym2}
\eeq 
and is generally a function of $r$ and $q$ although we will see later that for some qubit states, the $r$-dependence cancels out. We consider now two special cases. 
\medskip 

In the qubit case, where the ground state partition functions are just (\ref{funk}), (\ref{genZn}) reduces to
\beqa 
\mathcal{Z}_n^{k_1^{\epsilon_1}\ldots k_s^{\epsilon_s}}(r,q)
= \sum_{j_1=0}^{k_1}\sum_{j_2=0}^{k_2}\cdots \sum_{j_s=0}^{k_s}  \, \delta_{q,\sum_{\ell=1}^s j_\ell \epsilon_\ell} \left[\prod_{\ell=1}^s f_{j_\ell}^{k_\ell}(r) \right]^n\,.
\label{zn}
\eeqa
For a conformal field theory with $U(1)$ symmetry such as a massless complex free fermion or the compactified free boson  considered in \cite{Bonsignori:2019naz} the charged moments are proportional to the two-point function of a composite branch point twist field \cite{GS, Bonsignori:2019naz,Levi,BCDLR} of conformal dimension 
\beq 
\Delta_\alpha=\frac{c}{24}\left(n-\frac{1}{n}\right)+\frac{\tilde{\Delta}\alpha^2}{4n}\,,
\eeq 
where $\tilde{\Delta}$ is a number which depends on the type of theory (for example the compactification radius, if we are dealing with the compactified boson) and $c$ is the central charge. For unitary theories both $c$ and $\tilde{\Delta}$ are positive.
Therefore, the partition function is proportional to the Fourier transform of 
\beq 
\left( \frac{r}{\varepsilon}\right)^{-\frac{c}{6}(n-\frac{1}{n})-\frac{\tilde{\Delta}\alpha^2}{n}}\,,
\eeq 
where $\varepsilon$ is a UV cut-off and $r$ here represents the length of the region or distance between fields. Although the above is a very simple function of $\alpha$, the Fourier transform gives a non-trivial function of $q$ \cite{Bonsignori:2019naz}. The exact function can also be found in \cite{ourreview}, equation (24)\footnote{There is a typo in the normalisation of the Fourier transform in \cite{ourreview}. The should be no factor $\frac{1}{2\pi}$ since the Fourier variable is $2\pi\alpha$.}. It gives us the general formula for a CFT with $U(1)$ symmetry
\beqa 
\mathcal{Z}_n^{k_1^{\epsilon_1}\ldots k_s^{\epsilon_s}}(r,q)
= a_n(r) \sum_{j_1=0}^{k_1}\sum_{j_2=0}^{k_2}\cdots \sum_{j_s=0}^{k_s}  
e^{-\frac{\pi^2 (q-\sum_{\ell=1}^s j_\ell \epsilon_\ell)^2}{b_n(r)}} A^{j_1\ldots j_s}_n(r,q) \left[\prod_{\ell=1}^s f_{j_\ell}^{k_\ell}(r) \right]^n\,,
\label{znn}
\eeqa
with
\beq 
A^{j_1\ldots j_s}_n(r,q):={\rm{Erf}}\left(\frac{b_n(r)-2 \pi i (q-\sum_{\ell=1}^s j_\ell \epsilon_\ell)}{2\sqrt{b_n(r)}}\right)+ {\rm{Erf}}\left(\frac{b_n(r)+2 \pi i (q-\sum_{\ell=1}^s j_\ell \epsilon_\ell)}{2\sqrt{b_n(r)}}\right)\,,
\eeq
\beq 
a_n(r)=\left(\frac{\varepsilon}{r}\right)^{\frac{c}{6}\left(n-\frac{1}{n}\right)} 
\frac{\sqrt{\pi}} {2\sqrt{b_n(r)}}\qquad \mathrm{and}\qquad b_n(r)=\frac{\tilde{\Delta}}{n} \log\frac{r}{\varepsilon}\,,
\eeq 
and ${\rm Erf}(x)$ the error function. 

We can now employ (\ref{zn}) and (\ref{znn}) to obtain the symmetry resolved entropies. For a general qubit state we obtain 
\beqa 
S^{k_1^{\epsilon_1}\ldots k_s^{\epsilon_s}}(r,q)&=& \lim_{n\to 1} \frac{1}{1-n} \log \frac{\sum_{j_1=0}^{k_1}\sum_{j_2=0}^{k_2}\cdots \sum_{j_s=0}^{k_s}  \, \delta_{q,\sum_{\ell=1}^s j_\ell \epsilon_\ell} \left[\prod_{\ell=1}^s f_{j_\ell}^{k_\ell}(r) \right]^n}{\left[\sum_{j_1=0}^{k_1}\sum_{j_2=0}^{k_2}\cdots \sum_{j_s=0}^{k_s}  \, \delta_{q,\sum_{\ell=1}^s j_\ell \epsilon_\ell} \prod_{\ell=1}^s f_{j_\ell}^{k_\ell}(r)\right]^n} \nonumber\\
&=& - \frac{\sum_{j_1=0}^{k_1}\sum_{j_2=0}^{k_2}\cdots \sum_{j_s=0}^{k_s}  \, \delta_{q,\sum_{\ell=1}^s j_\ell \epsilon_\ell} \left[\prod_{\ell=1}^s f_{j_\ell}^{k_\ell}(r) \right]\log \left[\prod_{\ell=1}^s f_{j_\ell}^{k_\ell}(r)\right]}{\sum_{j_1=0}^{k_1}\sum_{j_2=0}^{k_2}\cdots \sum_{j_s=0}^{k_s}  \, \delta_{q,\sum_{\ell=1}^s j_\ell \epsilon_\ell} \left[\prod_{\ell=1}^s f_{j_\ell}^{k_\ell}(r) \right]} \nonumber\\
&+&\log \left(\sum_{j_1=0}^{k_1}\sum_{j_2=0}^{k_2}\cdots \sum_{j_s=0}^{k_s}  \, \delta_{q,\sum_{\ell=1}^s j_\ell \epsilon_\ell} \left[\prod_{\ell=1}^s f_{j_\ell}^{k_\ell}(r) \right]\right).
\label{moregen}
\eeqa
For a CFT we obtain instead
\small
\beqa
&&S^{k_1^{\epsilon_1}\ldots k_s^{\epsilon_s}}(r,q) = \frac{c}{3} \log \frac{r}{\varepsilon} -\frac{1}{2} + \log \mathcal{Z}_1^{k_1^{\epsilon_1}\ldots k_s^{\epsilon_s}}(r,q)\nonumber\\
&& -\frac{a_1(r) }{\mathcal{Z}_1^{k_1^{\epsilon_1}\ldots k_s^{\epsilon_s}}(r,q)} \sum_{j_1=0}^{k_1}\sum_{j_2=0}^{k_2}\cdots \sum_{j_s=0}^{k_s} e^{-\frac{\pi^2 (q-\sum_{\ell=1}^s j_\ell \epsilon_\ell)^2}{b_1(r)}} A^{j_1\ldots j_s}_1(r,q) \left[\prod_{\ell=1}^s f_{j_\ell}^{k_\ell}(r) \right]\log \left[\prod_{\ell=1}^s f_{j_\ell}^{k_\ell}(r)\right]\nonumber\\
&& +\frac{a_1(r) }{\mathcal{Z}_1^{k_1^{\epsilon_1}\ldots k_s^{\epsilon_s}}(r,q)} \sum_{j_1=0}^{k_1}\sum_{j_2=0}^{k_2}\cdots \sum_{j_s=0}^{k_s} \frac{\pi^2 (q-\sum_{\ell=1}^s j_\ell \epsilon_\ell)^2}{b_1(r)} \, e^{-\frac{\pi^2 (q-\sum_{\ell=1}^s j_\ell \epsilon_\ell)^2}{b_1(r)}}  A^{j_1\ldots j_s}_1(r,q) \left[\prod_{\ell=1}^s f_{j_\ell}^{k_\ell}(r) \right]\nonumber\\
&& +\frac{a_1(r) }{\mathcal{Z}_1^{k_1^{\epsilon_1}\ldots k_s^{\epsilon_s}}(r,q)} \sum_{j_1=0}^{k_1}\sum_{j_2=0}^{k_2}\cdots \sum_{j_s=0}^{k_s} \left[\prod_{\ell=1}^s f_{j_\ell}^{k_\ell}(r) \right] \sqrt{\frac{b_1(r)}{\pi}} \left(\frac{r}{\varepsilon}\right)^{-\frac{\tilde{\Delta}}{4}} (-1)^{q-\sum_{\ell=1}^s j_\ell \epsilon_\ell} .\nonumber\\
\eeqa
\normalsize
\subsection{Distinct Particles of Identical Charge: Qubit States}
Let us consider the case of qubits where all charges are identical, say $\epsilon_\ell= 1$ for all $\ell$ (choosing $\epsilon_\ell=-1$ does not change the end result). This case is particularly interesting because the formulae above admit certain simplifications that are not found in the general case. The only non-vanishing contribution to (\ref{zn}) corresponds to setting $q=\sum_\ell j_\ell$. This means that the non-vanishing terms in the multiple sum correspond to all partitions of the integer $q$ into $s$ non-negative and not necessarily distinct parts. We can do a bit more work to simplify this formula by using the explicit form of the functions $f_j^k(r)$ in terms of binomial coefficients and powers of $r$ and $1-r$. We then get
\beqa 
\mathcal{Z}_n^{k_1^{1}\ldots k_s^{1}}(r,q)
&=& \sum_{j_1=0}^{k_1}\sum_{j_2=0}^{k_2}\cdots \sum_{j_s=0}^{k_s}  \, \delta_{q,\sum_{\ell=1}^s j_\ell} \left[\prod_{\ell=1}^s {}_{k_\ell} C_{j_\ell}\right]^n \, r^{n\sum_{\ell=1}^s j_\ell} (1-r)^{n \sum_{\ell=1}^s (k_\ell-j_\ell)}\nonumber\\
&=&  r^{n q} (1-r)^{n (k-q)}  \sum_{j_1=0}^{k_1}\sum_{j_2=0}^{k_2}\cdots \sum_{j_s=0}^{k_s}  \, \delta_{q,\sum_{\ell=1}^s j_\ell} \left[\prod_{\ell=1}^s {}_{k_\ell} C_{j_\ell}\right]^n
\eeqa
where $k:=\sum_{\ell=1}^s k_\ell$. For $n=1$ using the generating function of the product of binomials, the constrained sum above can be easily computed to 
\beq 
\sum_{j_1=0}^{k_1}\sum_{j_2=0}^{k_2}\cdots \sum_{j_s=0}^{k_s}  \, \delta_{q,\sum_{\ell=1}^s j_\ell} \prod_{\ell=1}^s {}_{k_\ell} C_{j_\ell}={}_{k} C_{q}\,,
\label{29}
\eeq 
so that
\beq 
p(r,q):=\mathcal{Z}_1^{k_1^{1}\ldots k_s^{1}}(r,q)
= {}_{k} C_{q} r^{q} (1-r)^{k-q},
\eeq 
and we can then write a general formula for 
the R\'enyi entropies, namely
\beqa 
S_n^{k_1^1\ldots k_s^1}(q)&=&\frac{1}{1-n}\log{\frac{\mathcal{Z}_n^{k_1^{1}\ldots k_s^{1}}(r,q)}{(\mathcal{Z}_1^{k_1^{1}\ldots k_s^{1}}(r,q))^n}}\nonumber\\
&=&\frac{1}{1-n}\log{{\sum_{j_1=0}^{k_1}\sum_{j_2=0}^{k_2}\cdots \sum_{j_s=0}^{k_s}  \, \delta_{q,\sum_{i=1}^s j_i} \left[\frac{\prod_{i=1}^s {}_{k_i} C_{j_i}}{{\,}_k C_q}\right]^n}}\,. 
\label{snsn}
\eeqa 
The symmetry resolved von Neumann entropy can then be computed using
\beqa 
&& \lim_{n\rightarrow 1} \partial_n \mathcal{Z}_n^{k_1^{1}\ldots k_s^{1}}(r,q)
={}_{k} C_{q} \, r^q (1-r)^{k-q} \log\left[r^q (1-r)^{k-q}\right]\nonumber\\
&& \qquad + \, r^q (1-r)^{k-q} \sum_{j_1=0}^{k_1}\sum_{j_2=0}^{k_2}\cdots \sum_{j_s=0}^{k_s}  \, \delta_{q,\sum_{i=1}^s j_i} \left[\prod_{i=1}^s {}_{k_i} C_{j_i}\right] \log \left[\prod_{i=1}^s {}_{k_i} C_{j_i}\right]\,.
\eeqa 
and the definition (\ref{sym}) and (\ref{sym2}), giving
\beqa 
S^{k_1^1\ldots k_s^1}(q)\label{ss2}=- \sum_{j_1=0}^{k_1}\sum_{j_2=0}^{k_2}\cdots \sum_{j_s=0}^{k_s}  \, \delta_{q,\sum_{i=1}^s j_i} \left[\frac{\prod_{i=1}^s {}_{k_i} C_{j_i}}{{}_{k} C_{q}}\right] \log \left[\frac{\prod_{i=1}^s {}_{k_i} C_{j_i}}{{}_{k} C_{q}}\right]\,. \label{ss2}
\eeqa
Note that the special case of (\ref{ss2}) when all $k_i$ are 1 was presented in \cite{ourPartI}. 
From these results we can also obtain a closed formula for the number entropy  associated with such a qubit state, namely, from the definition
\beq 
n(r)=-\sum_q p(r,q)\log p(r,q)=-\sum_q  {}_{k} C_{q} r^{q} (1-r)^{k-q} \log\left[{}_{k} C_{q} r^{q} (1-r)^{k-q}\right]\,,
\label{number}
\eeq 
which is the Shannon entropy of the binomial distribution.

\begin{figure}[h!]
\begin{center}
	\includegraphics[width=7.8cm]{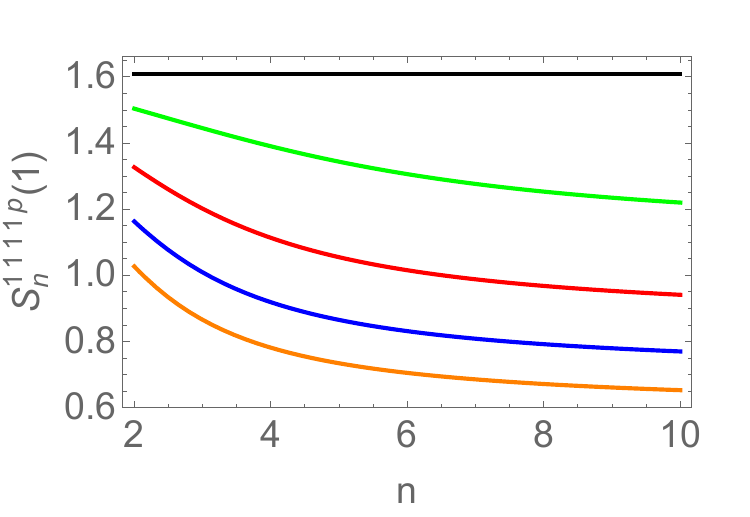}
	\includegraphics[width=7.8cm]{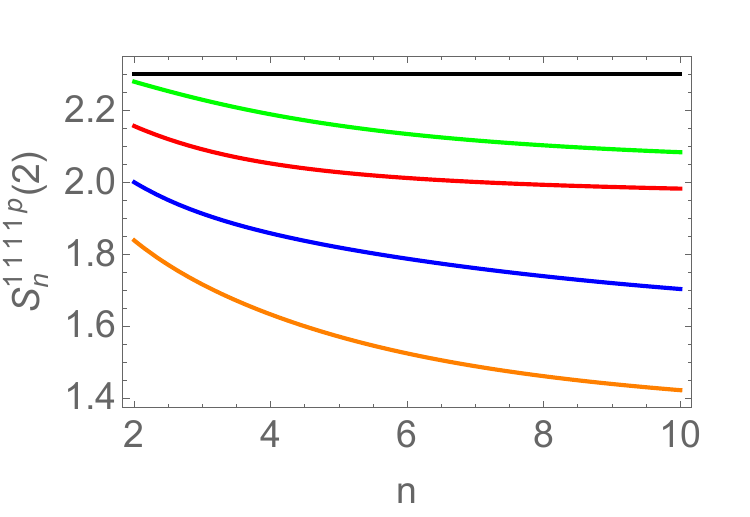}
    \includegraphics[width=7.8cm]{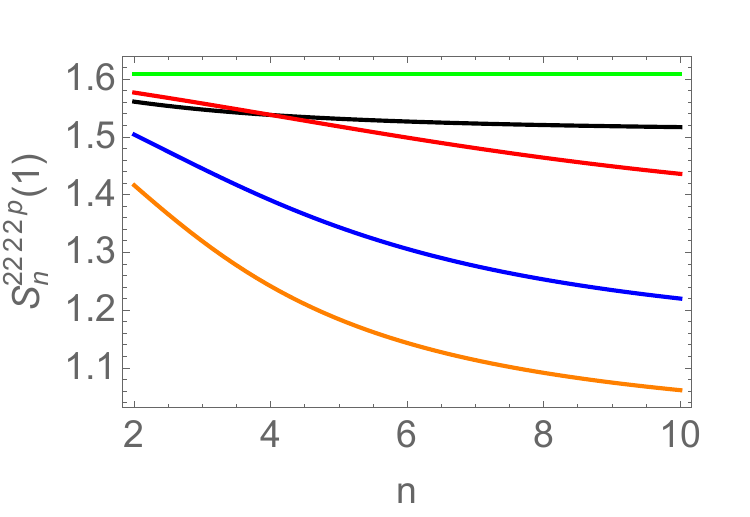}
	\includegraphics[width=7.8cm]{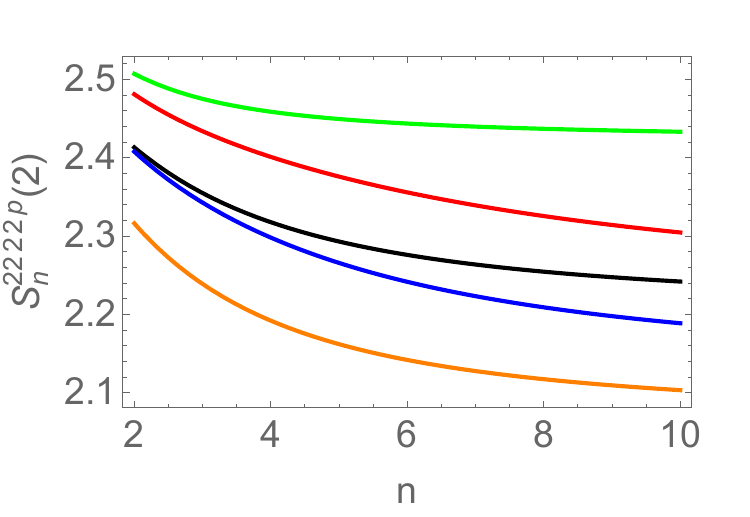}
				    \caption{The symmetry resolved R\'enyi entropies for charges $q=1 (\textrm{left}), 2 (\textrm{right})$ and various excited states consisting of five distinct groups of indistinguishable particles. In the first row the first four groups contain a single particle while in the second row they contain two. The fifth group has size  $p=1 (\textrm{black}), 2(\textrm{green}), 3(\textrm{red}), 4(\textrm{dark blue}), 5(\textrm{orange})$ excitations.  The entropy is highest when all groups consist of the same number of excitations and lowest for the largest value of $p$. Various values can be easily computed from the formulae (\ref{39}) and (\ref{32}). For instance, the black line in the top and bottom left figures stands at $\log 5$, which follows easily from the formulae (\ref{39}) and (\ref{32}). For the top right figure the black line stands at $\log 10$ which again follows from (\ref{32}). Finally, the top green curve in the bottom right figure is $n$-dependent but for $n$ large tends to the value $\log \frac{45}{4}$ which also follows from (\ref{32}). In these figures we have dropped the superindex `1' on the particle subgroups as they all have the same charge.}
				     \label{fig1}
    \end{center}
    \end{figure}

\begin{figure}[h!]
\begin{center}
	\includegraphics[width=7.8cm]{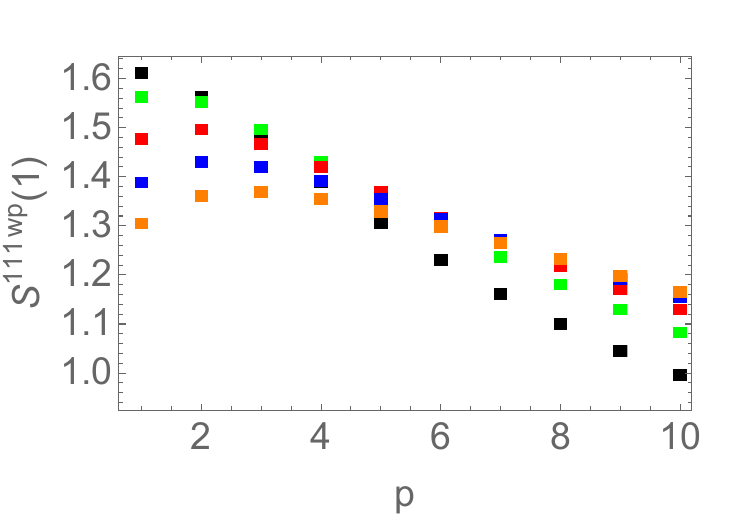}
	\includegraphics[width=7.8cm]{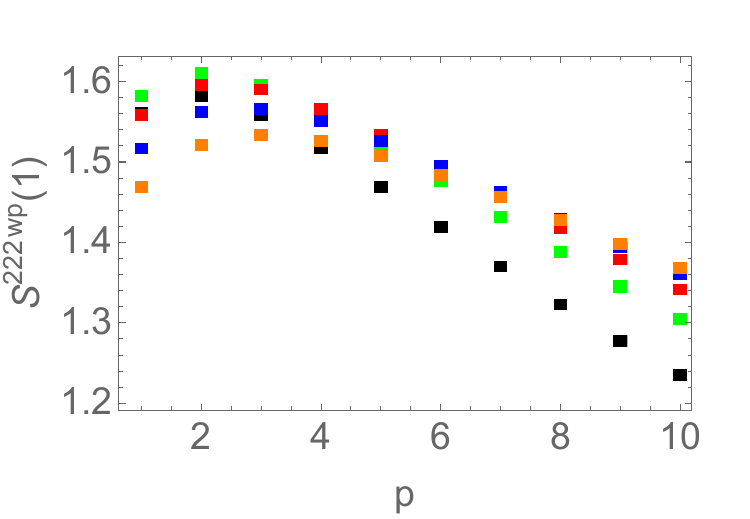}
    \includegraphics[width=7.8cm]{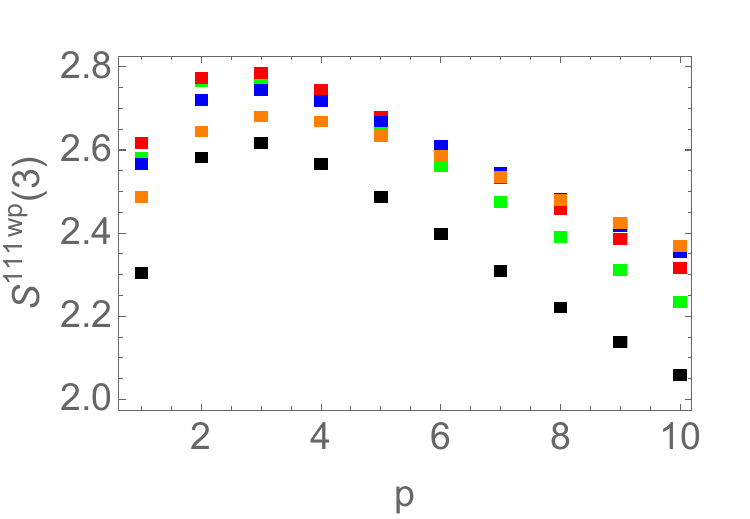}
	\includegraphics[width=7.8cm]{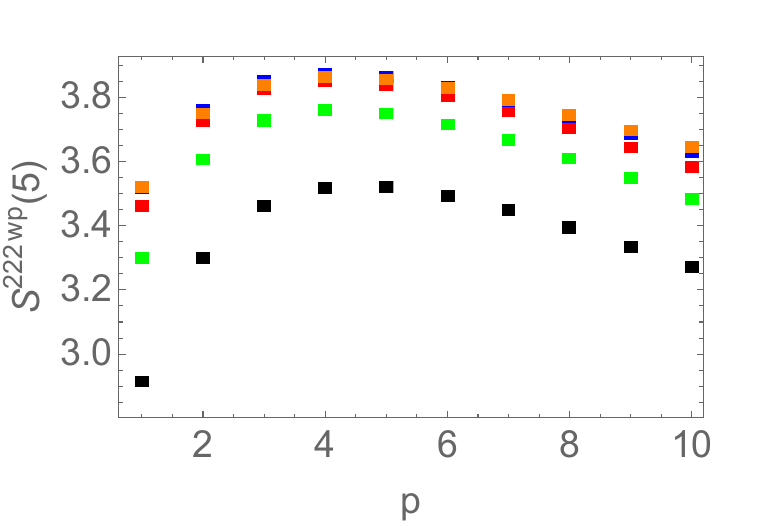}
				    \caption{The symmetry resolved von Neumann entropies for charges $q=1$ (top), $3$,
                 and $5$ (bottom) and various excited states consisting of five distinct groups of indistinguishable particles. In this example three groups consist of one (left)/two (right) excitation(s). The fourth group contains $w=1 (\textrm{black}), 2(\textrm{green}), 3(\textrm{red}), 4(\textrm{dark blue}), 5(\textrm{orange})$ excitations and $p$ varies from 1 to 10 as shown. We observe that the maximum value of the entropy is typically reached when $p=q=w$. In these figures we have dropped the superindex `1' on the particle subgroups as they all have the same charge.}
				     \label{fig2}
    \end{center}
    \end{figure}

    There are some interesting properties and special cases (ie $n=0,\infty$) of these formulae that we now discuss.
\begin{itemize}
\item {\bf $r$-Independence}: The SREEs, both R\'enyi and von Neumann, are $r$-independent, while the number and configuration entropies are not. In fact, all region size dependence of the total entropy is due to the probability function $p(r,q)$. This is a special feature of the case when all particles have the same charge and $q=\sum_i j_i$. Indeed, in \cite{ourPartI} we discussed some special cases when different charges are involved and showed that the results for the von Neumann and R\'enyi entropies are generally $r$-dependent, see also the next subsection. 
\item {\bf Vanishing SREEs}: The SREEs, both R\'enyi and von Neumann, are vanishing whenever $s=1$, that is for indistinguishable particles. In this case the full entropy coincides with the number entropy while the configuration entropy is vanishing.
\item {\bf Probabilistic Interpretation}: The formula (\ref{ss2}) is the Shannon entropy of all configurations of a certain total charge, with \beq 
\frac{\prod_{i=1}^s {}_{k_i} C_{j_i}}{{}_{k} C_{q}}\,,
\label{prob}
\eeq 
representing the probability of a configuration consisting of $s$ subsets of $k_1,\ldots ,k_s$ excitations each, of which $j_1,\ldots, j_s$ are in region $A$ and the rest in region $\bar{A}$. As we know, the Shannon entropy, is maximised when all probabilities are identical. In our case the choice of $j_i$ is restricted by the choice of $q$ of $k_i$ so determining which configuration maximises the entropy is non-trivial, see Figs.~1 and 2.
\item {\bf Symmetry Resolved Single-Copy Entropy}: Given a set of fixed values $k_1,\ldots, k_s$ and a total charge $q$, there exists at least one configuration $j_1^m,\ldots, j^m_s$ which maximises the probability (\ref{prob}). If we know this configuration, then the single-copy symmetry resolved entropy can be defined as, 
\beq 
S_\infty^{k_1^1\ldots k_s^1}(q):=\lim_{n\rightarrow \infty} S_n^{k_1^1\ldots k_s^1}(q)\,,
\eeq 
by generalising the standard definition \cite{Single1,Single2,Single3,Single4}. In our case, it is given by
\beq 
S_\infty^{k_1^1\ldots k_s^1}(q)=-\log  \left[\frac{\prod_{i=1}^s {}_{k_i} C_{j_i^m}}{{}_{k} C_{q}}\right]\,.
\eeq 
\item {\bf Symmetry Resolved R\'enyi-0 Entropy}: Another special case of the formula (\ref{snsn}) is the symmetry resolved generalisation of what has been termed R\'enyi-0 entropy in \cite{Agon}, that is:
\beq 
S_0^{k_1^1\ldots k_s^1}(q)=:=\lim_{n\rightarrow 0} S_n^{k_1^1\ldots k_s^1}(q)\,.
\eeq
This quantity has been related to the effective entanglement temperature in the entanglement hamiltonian. In our case 
\beq
S_0^{k_1^1\ldots k_s^1}(q)=\log \mathcal{R}^{k_1\ldots k_s}(q)\,,
\eeq 
where $\mathcal{R}^{k_1\ldots k_s}(q)$ is the number of partitions of $q$ into $s$ non-negative, not necessarily distinct parts with part $i$ less or equal than the value $k_i$. 
\end{itemize}
Some special cases of these formulae were studied in \cite{ourPartI}, for instance the case when all $k^1_i=1$. There are other cases of fixed charge, where formulae simplify considerably. For example the $q=0$ sector gives
$S_n^{k_1^1\ldots k_s^1}(0)=
S^{k_1^1\ldots k_s^1}(0)= 0$
which just indicates the obvious fact that we cannot have a charge zero sector containing any particles if all particles have the same charge. Less trivially, for $q=1,2,3$ we can write
\beq 
S_n^{k_1^1\ldots k_s^1}(1)= \frac{1}{1-n} \log\sum_{i=1}^s \left(\frac{k_i}{k}\right)^n,\qquad 
S^{k_1^1\ldots k_s^1}(1)=- \sum_{i=1}^s \frac{k_i}{k} \log \frac{k_i}{k}\,,
\label{39}
\eeq 
\beqa 
S_n^{k_1^1\ldots k_s^1}(2) &=& \frac{1}{1-n}\log\left[\sum_{\substack{i=1\\ k_i\geq 2}}^s \left(\frac{k_i(k_i-1)}{k(k-1)}\right)^n + \sum_{1\leq i<j\leq s} \left(\frac{2 k_i k_j}{k(k-1)}\right)^n\right]\,,
\label{32}
\eeqa 
\beqa 
S^{k_1^1\ldots k_s^1}(2) = -  \sum_{\substack{i=1\\ k_i\geq 2}}^s \frac{k_i(k_i-1)}{k(k-1)}  \log \left[\frac{k_i(k_i-1)}{k(k-1)}\right] -  \sum_{1 \leq i<j\leq s}  \frac{2 k_i k_j}{k(k-1)} \log\left[\frac{2k_i k_j}{k(k-1)}\right]\,,
\label{33}
\eeqa 
and 
\beqa 
S_n^{k_1^1\ldots k_s^1}(3) = \frac{1}{1-n}\log\left[\sum_{\substack{i=1\\ k_i\geq 3}}^s \left[\frac{{}_{k_i} C_3}{{}_{k} C_3}\right]^n + \sum_{\substack{1 \leq i<j\leq s\\ k_j\geq 2}} \left(k_i \frac{{}_{k_j}C_2}{{}_{k} C_3}\right)^n + \sum_{1\leq i<j<r\leq s} \left(\frac{k_i k_j k_r}{{}_{k} C_3}\right)^n\right]\,,
\eeqa 
\beqa 
S^{k_1^1\ldots k_s^1}(3) &=& -  \sum_{\substack{i=1\\ k_i\geq 3}}^s \frac{{\,}_{k_i} C_3}{{\,}_k C_3}  \log \left[\frac{{\,}_{k_i} C_3}{{\,}_k C_3}\right] - \sum_{1 \leq i<j<r\leq s}  \frac{k_i k_j k_r}{{\,}_k C_3} \log\left[\frac{k_i k_j k_r}{{\,}_k C_3}\right] \nonumber\\
&& -\sum_{\substack{1 \leq i<j\leq s\\ k_i\geq 2}}  \frac{ k_j {\,}_{k_i} C_2}{ {\,}_k C_3} \log\frac{ k_j {\,}_{k_i} C_2}{ {\,}_k C_3}\,,
\eeqa 
and so on. Note that the conditions $k_i \geq 2, 3$ above stem from the fact that $k_i$ must always be greater or equal than $j_i$. Thus the sum over $i$ with restricted $k_i$ means that only those terms involving $k_i$ satisfying the given condition are included in the sum. For example, it is only possible to have two identical particles from the set $k_i$ in a given region, if $k_i\geq 2$ in the first place. These formulae have been employed in order to obtain Figs.~\ref{fig1} and \ref{fig2}.

    \subsection{Distinct Particles of Distinct Charges: Qubit States}
    \label{onedis}
    We would like to  close our discussion by considering another general case, namely the case of $s$ groups of $k_i$ identical excitations each, where one of the groups, say the last one, contains $k_s$ particles of negative charge, while all other groups contain particles of positive charge. In this section, we will use the superindices $\pm$ to indicate positive/negative charge of the excitations. This is a generalisation of special cases we discussed in \cite{ourPartI} and allows us to show more clearly why the $r$-independence observed in the previous subsection is a special feature of the case of identical charge. We consider again the general formula (\ref{zn}) and adapt it to our case as
    \beqa 
   \mathcal{Z}^{k_1^{+}\ldots k_{s-1}^+ k_s^{-}}_n(r,q)={\sum_{j_1=0}^{k_1}\sum_{j_2=0}^{k_2}\cdots \sum_{j_s=0}^{k_s}  \, \delta_{q+j_s,\sum_{\ell=1}^{s-1} j_\ell} \left[\prod_{\ell=1}^s f_{j_\ell}^{k_\ell}(r) \right]^n}
   \eeqa 
   and employing the $\delta$ function and the definition of $f_j^k(r)$ it is possible to carry out the product over the $r$-dependent terms giving,
   \beq 
\mathcal{Z}^{k_1^{+}\ldots k_{s-1}^+ k_s^{-}}_n(r,q)= r^{n q} (1-r)^{n(k-q)} \sum_{j_1=0}^{k_1}\sum_{j_2=0}^{k_2}\cdots \sum_{j_s=0}^{k_s}  \, \delta_{q+ j_s,\sum_{\ell=1}^{s-1} j_\ell } \left(\frac{r}{1-r}\right)^{2 n j_s}\left[\prod\limits_{\ell=1}^s  {}_{k_\ell} C_{j_\ell}\right]^n\,.
   \eeq 
For $n=1$, we can employ a similar result to (\ref{29}) and simplify the formula by carrying out the first $s-1$ sums to 
\beqa 
\mathcal{Z}^{k_1^{+}\ldots k_{s-1}^+ k_s^{-}}_1(r,q)&=& r^{q} (1-r)^{(k-q)} \sum_{j_s=0}^{k_s} \left(\frac{r}{1-r}\right)^{2 j_s} \left[{}_{k_s} C_{j_s}\right] \left[{}_{k-k_s} C_{q+j_s}\right]\nonumber\\
&=& r^{q} (1-r)^{(k-q)}  \left[{}_{k-k_s} C_{q}\right] \,{}_{2} F_1 \left(-k_s,q+k_s-k,1+q; \left(\frac{r}{1-r}\right)^{2}\right)\,,
   \eeqa 
   in terms of a hypergeometric function. This is the probability $p(r,q)$ of measuring a charge $q$ and can be plugged into the usual formula to obtain the number entropy as in (\ref{number}). This reduces to polynomials in powers of $r$ and $1-r$ for specific integer values of $k, k_s$ and $q$. For example, for $k_s=1$:
\beqa 
\mathcal{Z}^{k_1^{+}\ldots k_{s-1}^+ 1^{-}}_1(r,q)= r^{q} (1-r)^{(k-q)}  \left[{}_{k-1} C_{q}\right] \left[1+\left(\frac{k}{1+q}-1\right)\left(\frac{r}{1-r}\right)^{2}\right]\,.
   \eeqa 
In order to compute the entropy, it is useful to compute
\beqa 
&& \frac{\partial_n \mathcal{Z}^{k_1^{+}\ldots k_{s-1}^+ k_s^{-}}_n(r,q)}{\mathcal{Z}^{k_1^{+}\ldots k_{s-1}^+ k_s^{-}}_n(r,q)} = \log\left[r^{q}(1-r)^{k-q}\right] \\
&& + \frac{\sum_{j_1=0}^{k_1}\sum_{j_2=0}^{k_2}\cdots \sum_{j_s=0}^{k_s}  \, \delta_{q+ j_s,\sum_{\ell=1}^{s-1} j_\ell } \left(\frac{r}{1-r}\right)^{2 n j_s}\left[\prod\limits_{\ell=1}^s  {}_{k_\ell} C_{j_\ell}\right]^n  \log\left[\left(\frac{r}{1-r}\right)^{2 j_s}\left[\prod\limits_{\ell=1}^s  {}_{k_\ell} C_{j_\ell}\right]\right]}{\sum_{j_1=0}^{k_1}\sum_{j_2=0}^{k_2}\cdots \sum_{j_s=0}^{k_s}  \, \delta_{q+ j_s,\sum_{\ell=1}^{s-1} j_\ell } \left(\frac{r}{1-r}\right)^{2 n j_s}\left[\prod\limits_{\ell=1}^s  {}_{k_\ell} C_{j_\ell}\right]^n}\nonumber
\eeqa 
\begin{figure}[h!]
    \centering
     \includegraphics[width=0.45\linewidth]{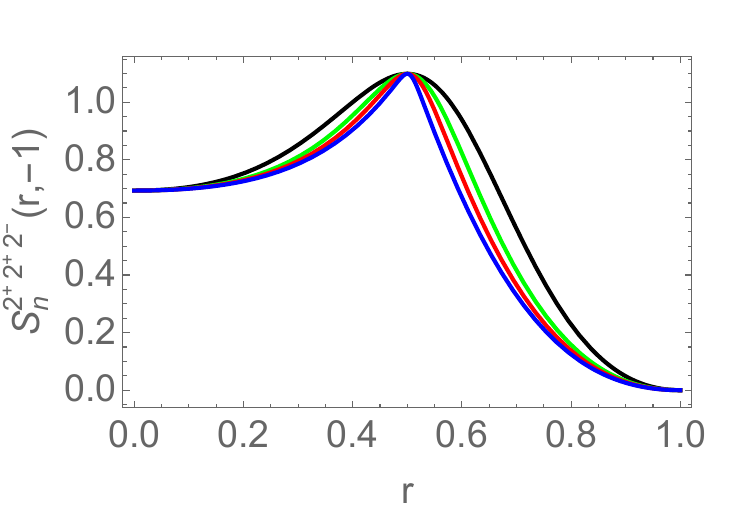}
    \includegraphics[width=0.45\linewidth]{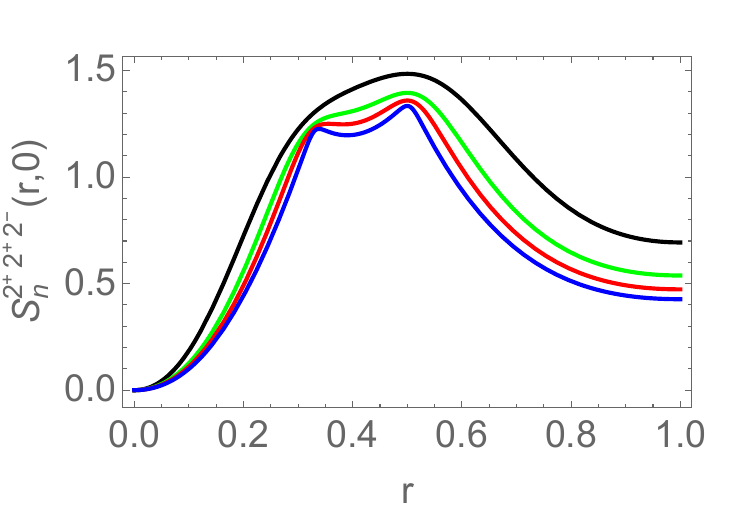}
    \includegraphics[width=0.45\linewidth]{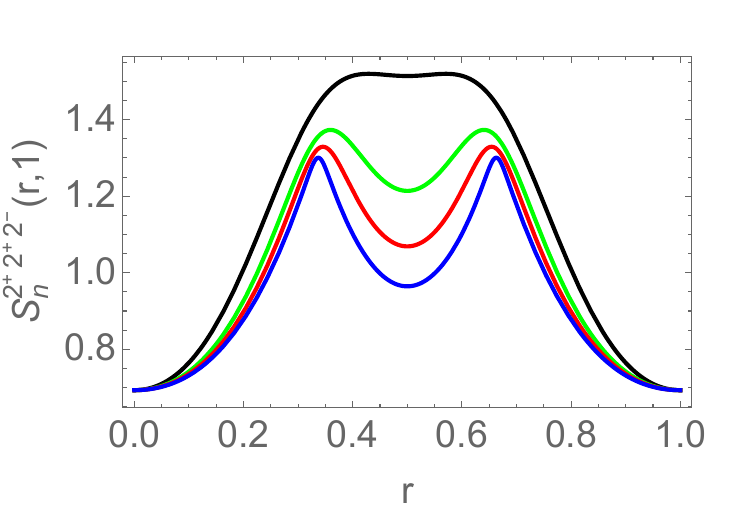}
    \includegraphics[width=0.45\linewidth]{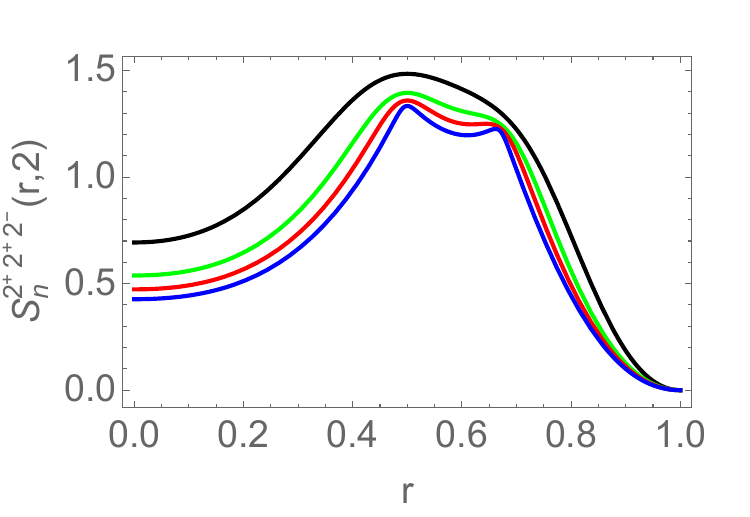}
    \caption{The symmetry resolved R\'enyi entropies $S_n^{2^+ 2^+ 2^-}(r,q)$ for charges $q=-1,0,1,2$ of a state consisting of three groups of two identical particles each. Two of the groups have positive charge and one group has negative charge. The different colours correspond to different $n$. The values are $n=2 $(black), $n=4$(green), $n=7$ (red) and $n=20$ (blue). The entropies exhibit the property $S_n^{2^+ 2^+ 2^-}(r,q)=S_n^{2^+ 2^+ 2^-}(1-r, 2-q)$ which is state-dependent.}
    \label{newfig}
\end{figure}
   Then, with the usual definitions we obtain
   \beqa
    S^{k_1^{+}\ldots k_{s-1}^+ k_s^{-}}_n(r,q) = \frac{1}{1-n} \log \frac{\sum_{j_1=0}^{k_1}\sum_{j_2=0}^{k_2}\cdots \sum_{j_s=0}^{k_s}  \, \delta_{q+ j_s,\sum_{\ell=1}^{s-1} j_\ell } \left(\frac{r}{1-r}\right)^{2n j_s}\left[\prod\limits_{\ell=1}^s  {}_{k_\ell} C_{j_\ell}\right]^n}{\left[\left[{}_{k-k_s} C_{q}\right] \,{}_{2} F_1 \left(-k_s,q+k_s-k,1+q; \left(\frac{r}{1-r}\right)^{2}\right) \right]^n}\,,
\eeqa
and 
 \beqa
    && S^{k_1^{+}\ldots k_{s-1}^+ k_s^{-}}(r,q) =\log\left[\left[{}_{k-k_s} C_{q}\right] \,{}_{2} F_1 \left(-k_s,q+k_s-k,1+q; \left(\frac{r}{1-r}\right)^{2}\right) \right] \\
    &&  -\frac{\sum_{j_1=0}^{k_1}\sum_{j_2=0}^{k_2}\cdots \sum_{j_s=0}^{k_s}  \, \delta_{q+ j_s,\sum_{\ell=1}^{s-1} j_\ell } \left(\frac{r}{1-r}\right)^{2 j_s}\left[\prod\limits_{\ell=1}^s  {}_{k_\ell} C_{j_\ell}\right]  \log\left[\left(\frac{r}{1-r}\right)^{2 j_s}\left[\prod\limits_{\ell=1}^s  {}_{k_\ell} C_{j_\ell})\right]\right]}{\left[\left[{}_{k-k_s} C_{q}\right] \,{}_{2} F_1 \left(-k_s,q+k_s-k,1+q; \left(\frac{r}{1-r}\right)^{2}\right) \right]}\,. \nonumber
\eeqa 
Therefore, while the von Neumann entropy has contributions that are $r$-independent, it is not independent of $r$ as we see also in Fig.~\ref{newfig}.
\section{Conclusions}
In this work we have returned to our study of the symmetry resolved entanglement entropy of excited states following on the steps of previous works \cite{ourPartI,ourPartII,ourPartIII}. We have obtained exact formulae for the von Neumann and R\'enyi entropies in cases where the ground state partition function is known exactly. We considered generic excited states in the presence of $U(1)$ symmetry, for the case when the theory is a CFT and when the state is written in a qubit basis. 

For qubit states, when all excitations have the same charge, there are significant simplifications and we obtain \ref{snsn} and \ref{ss2}. In particular, we find that the SREEs are independent of region size, while the configurational and number entropies are not.  In fact, all dependence on region size enters through the probability function $p(r,q)$. In more general cases, such as the case considered in subsection (\ref{onedis}) we find that all entropies depend on region size. 

An interesting open problem is to extend this formalism to other continuous symmetries, especially non-abelian ones such as $SU(N)$. Such work could provide a framework for connecting algebraic properties with physical phenomena such as phase transitions, topology and universality. Such connections play a crucial role in areas such as quantum computing, topological materials and quantum field theory.

\label{conclude}

\medskip

\noindent {\bf Acknowledgments:} This paper is part of a series of works \cite{ourPartI,ourPartII,ourPartIII} on symmetry resolved entanglement measures for excited states, involving several collaborators. We would like to thank our collaborators in previous work, Luca Capizzi, Cecilia De Fazio and Michele Mazzoni. Luc\'ia Santamar\'ia-Sanz is grateful to the Spanish Ministry of Science and Innovation (MICIN), and the Regional Government of Castilla y Le\'on (Junta de Castilla y Le\'on), for the support through the European Union NextGenerationEU funds (PRTRC17.I1). This work has also been supported by the grant PID2023-148373NB-I00 funded by MCIN/AEI/ 10.13039/501100011033/FEDER, UE. Luc\'ia Santamar\'ia-Sanz also thanks the Department of Mathematics of City St George's, University of London, for hospitality during a two-month visit in June-July 2024 when this project was initiated. 

\bibliography{Ref}
\end{document}